\documentclass[12pt]{article}
\usepackage{graphicx,amsmath}
\usepackage{units}
\usepackage{color}

\newlength{\dinwidth}
\newlength{\dinmargin}
\def\lapproxeq{\lower .7ex\hbox{$\;\stackrel{\textstyle                                                    
<}{\sim}\;$}}                                                    
\def\gapproxeq{\lower .7ex\hbox{$\;\stackrel{\textstyle                                                    
>}{\sim}\;$}}                                                    
\def\be{\begin{equation}}                                                    
\def\ee{\end{equation}}                                                    
\def\bea{\begin{eqnarray}}                                                    
\def\eea{\end{eqnarray}}

\def\GeV{\rm GeV}

\def\sh{\hat s}
\def\sh2{{\hat s}^2}

\begin{document}

\begin{flushright}                                                    
IPPP/14/85  \\
DCPT/14/170 \\                                                    
\today \\                                                    
\end{flushright} 

\vspace*{1.0cm}

\begin{center}
{\Large \bf Partonic transverse momenta in soft collisions}\\

\vspace*{1cm}
                                                   
V.A. Khoze$^{a,b}$, A.D. Martin$^a$ and M.G. Ryskin$^{a,b}$  \\                                                    
                                                   
\vspace*{1.0cm}                                                    
$^a$ Institute for Particle Physics Phenomenology, University of Durham, Durham, DH1 3LE \\                                                   
$^b$ Petersburg Nuclear Physics Institute, NRC Kurchatov Institute, Gatchina, St.~Petersburg, 188300, Russia \\          
                                                    
\vspace*{1cm}

\begin{abstract} 

The partonic transverse momentum, $k_t$ distribution plays a crucial role in driving high-energy hadron interactions. If $k_t$ is limited we have old fashioned Regge Theory. If $k_t$ increases with energy the interaction may be described by perturbative QCD. We use
BFKL diffusion in $\ln k_t$, supplemented by a stronger absorption of low $k_t$ partons, to estimate the growth of the mean transverse momenta $\langle k_t\rangle$ with energy.
This growth reveals itself in the distribution of secondaries produced at the collider energies. We present a simple, BFKL-based, model to demonstrate the possible size of the effect. Moreover, we propose a way to evaluate experimentally the shape of the parton transverse momenta distribution by studying the spectra of the ($D$ or $B$) mesons which contain one heavy quark.
\end{abstract}                                                        
\vspace*{0.5cm}                                                    
                                                    
\end{center}

\section{Introduction  \label{sec:1}} 
Contrary to old Regge theory, where it was {\em assumed} that the transverse
momenta of all the particles are limited, QCD is a logarithmic theory where there is a possibility that the parton's (quark, gluon) transverse momenta may increase during the evolution. In particular, already in leading order (LO) BFKL evolution there is  diffusion in $\ln k_t$ space~\cite{Lip86}. From the experimental point of view, it is relevant to note that the growth of the mean transverse momenta, $\langle p_t\rangle$, of secondary hadrons with collider energy was observed at the Tevatron and at the LHC (see e.g.~\cite{pT}). 
In order to describe this growth in DGLAP-based Monte Carlo generators \cite{Pythia,DGLAPMC} an additional 
infrared cutoff, $k_{\rm min}$, was introduced.
Of course, in any case, we need a  cutoff to avoid the infrared divergence of the amplitude of the hard (parton-parton interaction) subprocess. However, at first sight, we would expect that this cutoff to have its origin in confinement. It should be less than 1 GeV and should not depend on energy. On the contrary, it turns out, that to
reproduce the energy dependence of the data, the value of $k_{\rm min}$
should increase as $k_{\rm min}\propto s^{0.12}$~\cite{Pythia}; such that at the Tevatron energy $k_{\rm min}\simeq 2$ GeV, while at the LHC $k_{\rm min}\simeq 3$ GeV.

In Section 2  we present a simple model which accounts for BFKL $\ln k_t$ diffusion, together with the absorptive effects which additionally suppress the low $k_t$ partons, since the absorptive cross section behaves as $\sigma^{\rm abs}\propto 1/k^2_t$. That is, we now have a {\it dynamically} induced infrared cutoff. In Section 3 we use this model to obtain the expected energy and rapidity dependence of $k_t$ distributions.  In Section 4, we discuss the possibility to directly study these effects experimentally by measuring the $p_t$ spectra
of $D$ (and/or $B$) mesons. Due to the strong leading particle effect
(see e.g \cite{LPE}) the transverse momentum of mesons which contain a heavy quark is close to the transverse momentum of the heavy quark. Moreover, final state interactions and confinement do not appreciably distort the original distribution of these  heavy mesons.

\section{BFKL-based model}
\begin{figure} 
\begin{center}
\vspace{-3.5cm}
\includegraphics[height=10cm]{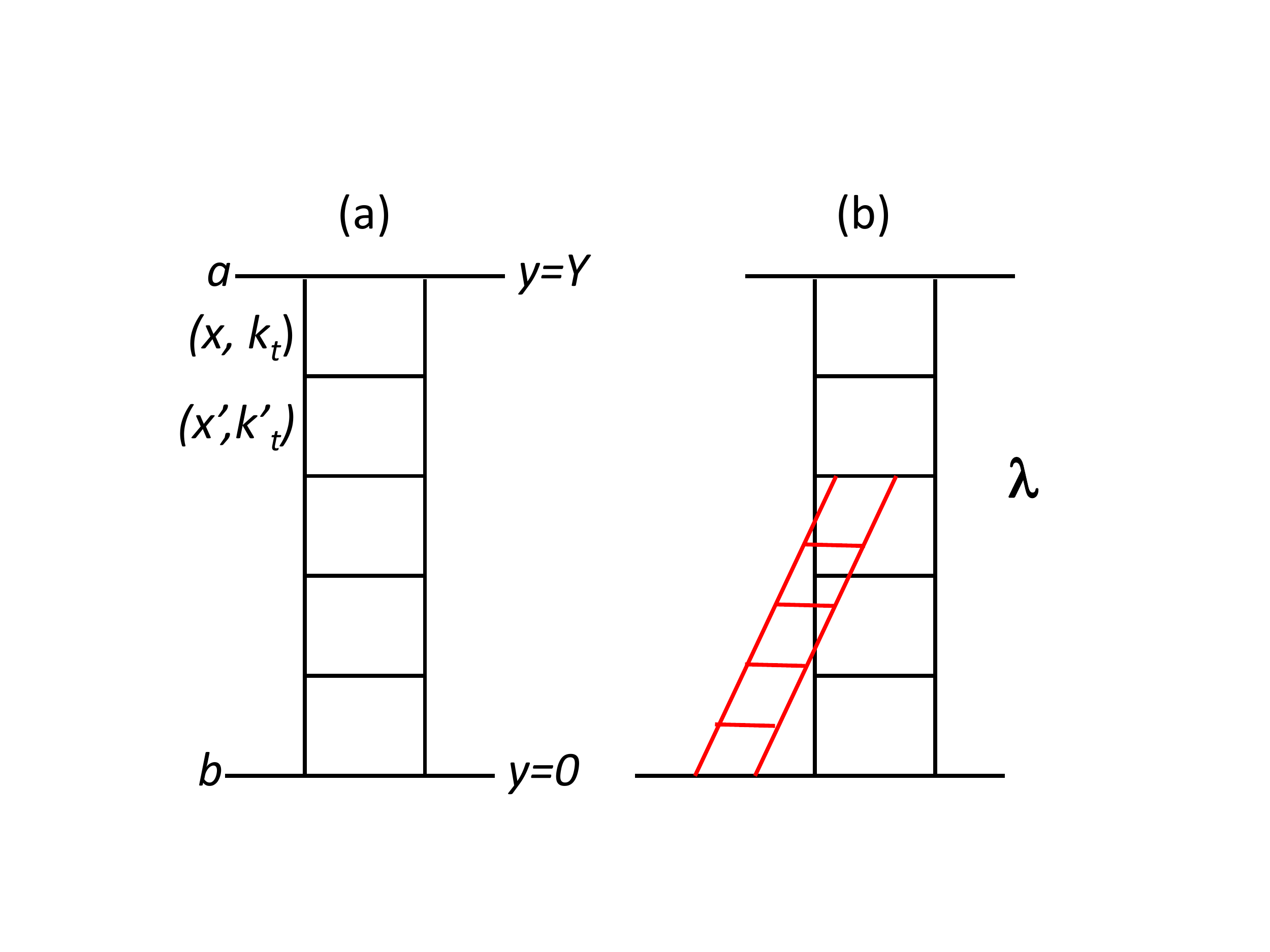}
\vspace{-1.5cm}
\caption{\sf (a) BFKL ladder diagram; (b) the ladder structure of the triple-pomeron coupling.}
\label{fig:f1}
\end{center}
\end{figure}

The original BFKL equation~\cite{BFKL} may be written as an integral equation for the gluon distribution unintegrated over $k_t$,
\be
f(x,k_t)=\partial[xg(x,k^2_t)]/\partial [d\ln k^2_t],
\ee   
in the form:
\be
\label{eq:uBFKL}
f(x,k_t)=f_0(x,k_t) + \frac{\alpha_s}{2\pi}  \int_x^1 \frac{d z}{z} \int_{k_0}^\infty \frac{d^2k'_t}{\pi} ~{\cal K}(k_t,k'_t,z)~f(x/z,k'_t),
\ee
where the kernel is evaluated as
\be
\label{eq:kernel}
{\cal K}(k_t,k_t',z)f(x/z,k'_t)=
2N_c\frac{k^2_t}{k^{'2}_t}\left[\frac{f(x/z,k'_t)-f(x/z,k_t)}{|k^{'2}_t-k^2_t|}
~+~\frac{f(x/z,k_t)}{\sqrt{4k^{'4}_t+k^4_t}}\right]~.
\ee
The first term 
 in the kernel\footnote{Here we have already integrated over the azimuthal angle $\phi$ assuming, similar to the DGLAP case, a flat $\phi$ dependence of $f$; that is, we consider the zero harmonic, which corresponds to the trajectory with the rightmost intercept.} can be understood as the effect of the emission of a daughter gluon with momentum ($x,k_t$) from a parent gluon with momentum ($x'=x/z, k_t'$). This generates the ladder structure of the pomeron sketched in Fig. \ref{fig:f1}(a). The remaining two terms in the kernel (depending on $f(x/z,k_t)$) account for the loop corrections which occur in the trajectory of $t$-channel reggeized gluons. 

 It is important to evolve in $k_t$ (as well as $x$) to be able to understand the origin and the behaviour of the dynamical infrared cutoff -- that is to see how $k_t(s,y)$ distribution is generated within perturbative QCD. Here $y=$ ln$(1/x)$ is the rapidity of the parton.  This dynamically generated cutoff affects (i) the $p_T$ distribution of secondary hadrons, (ii) the slope, $\alpha'_P$, of the (QCD) pomeron trajectory, and (iii) the values of the triple- and multi-pomeron couplings which control the predictions of the cross sections for diffractive dissociation.
 
Equation (\ref{eq:uBFKL}) can be solved numerically, starting, for example, from an input gluon with 
\be
\label{4}
f_0(x,k_t)=\alpha_s(k_t)\delta(x-x_0),
\ee
where we take $x_0=0.2$. Since the probability to have a large $k_t$ gluon should be  suppressed by the small QCD coupling, we have included in (\ref{4}) the factor $\alpha_s(k_t)$ . We use the one-loop running coupling
$\alpha_s(k_t^2)$ with $\Lambda_{\rm QCD}=0.15$ GeV, and the number of light quarks to be $n_f=4$. Besides this, we account for the
simple kinematical constraint -- when the parton carries fraction $x$ of the initial proton momenta, the transverse momentum $k_t$ cannot exceed the value $k_{t,\rm max}=\sqrt{xs}$, where $\sqrt{s}$ is the initial energy.  

It is natural to approximate the input by taking $x_0=0.2$. The reasons are as follows. For BFKL evolution we have to consider small $x$, but we would like to cover the largest possible rapidity interval. Therefore we start with $x_0=0.2$, reserving a larger $x$ interval for the valence quarks, and possible Good-Walker diffractive eigenstates \cite{GW}, which describe low-mass diffractive dissociation. Moreover the typical DGLAP input gluon has a $(1-x)^5$ type distribution corresponding to a mean of $x$ of about 0.2.

To include the effects of absorption (that is the rescattering of intermediate partons along the ladder) we follow~\cite{hs} and multiply the BFKL kernel ${\cal K}$ of eq.  (\ref{eq:kernel}) by a canonical absorptive factor of the form $\exp(-\lambda\Omega(y,k_t)/2)$ which depends on the rapidity, $y=\ln(1/x)$, and the $k_t$ of the current parton.  Here $\Omega$ is the 
optical density of the target gluon, while the factor $\lambda$ accounts for the value of the triple-pomeron vertex, such that $\lambda\Omega$ is the opacity of an incoming proton - `current' parton interaction\footnote{Recall that, in the eikonal framework, exp$(-\Omega)$ is the probability of no inelastic interaction.  
Since we consider the amplitude, and not the cross section, we put $\Omega /2$ in (\ref{abs}), rather than the full opacity $\Omega$.}. However, we must account for the absorption by both the incoming beam $(a)$ and the target $(b)$ protons interacting with intermediate partons. That is actually the absorptive factor reads 
\be
\label{abs}
S=\exp(-\lambda[\Omega^b(y,k_t)+\Omega^a(y',k_t)]/2),
\ee
where $y$ ($y'$) is the rapidity difference between the beam (target) proton and the current, intermediate gluon in the BFKL evolution. Denoting the rapidity separation between the beam and the target protons by $Y$, we have $y'=Y-y$.   

The simplest absorptive effect comes from the triple-pomeron diagram shown in Fig. \ref{fig:f1}(b).  As in~\cite{hs}, we use the Leading Log expression for the BFKL triple-pomeron vertex, that is~\cite{GLR,BW,Bal}
\be
\label{lambda}
\lambda=N_c\alpha_S(k_t)\Theta(k'_t-k_t).
\ee 
The $\Theta$-function reflects the fact that (after averaging over the azimuthal angle)
the large-size pomeron (i.e. the ladder with small $k'_t$) does not `see' the small-size colourless object described by the BFKL pomeron component with $k_t>k'_t$.

Note that the suppression factor, written in the form (\ref{abs}), includes not just the triple-pomeron diagram, but also a series of the multi-pomeron contributions generated by the vertices, $g^n_m$, coupling $n$ to $m$ pomerons. Here we prefer to take the simple eikonal-like expression
\be
g^n_m=\Omega(\lambda\Omega)^{n+m-2}
\ee 
which satisfies the AGK cutting rules~\cite{AGK}. However this means that we have to replace the exponents $\exp(-\lambda\Omega/2)$ in (\ref{abs}) by the factor
\be
\label{abs1}
\frac{1-\exp(-\lambda\Omega)}{\lambda\Omega}.
\ee
So now the absorptive factor (\ref{abs}) becomes
\be
\label{abs-S}
S(y,y',k_t,k'_t)=\frac{[1-\exp(-\lambda\Omega^b(y,k_t))]}{\lambda\Omega^b(y,k_t)}
\frac{[1-\exp(-\lambda\Omega^a(y',k_t))]}{\lambda\Omega^a(y',k_t)}
\ee
with the $\lambda(k'_t,k_t)$ given by (\ref{lambda}).


In terms of gluon density $f(x,k_t)$, the `differential' opacity $\Omega^b(y,k_t)$ of hadron $b$ (corresponding to the contribution from the $d\ln(k^2_t)$ interval) reads\footnote{This equation follows after integrating eq.(17) of \cite{hs} over the impact parameter, $b$, or from \cite{BW}.}
\be
\label{opa}
\lambda\Omega^b(y,k'_t)=N_c\pi^2\alpha_s(k'_t)
\frac{f^b(y,k'_t)}{16\pi k^{'2}_t\ B_g}\ ,
\ee
where $B_g/2$ is the $t$-slope of the initial "constituent gluon" form factor; we take\footnote{There are several arguments in favour of the effective slope $B_g$ being of the order of 1 $\GeV^{-2}$; that is in favour of the small size `hot spot' transverse area occupied by our gluon amplitude. The first reason, is the small radius of the gluonic form factor of the proton calculated using QCD sum rules \cite{QCDsr}. The next argument is the small value of the effective slope of the pomeron trajectory observed experimentally. Further evidence is the success of the additive quark model, $\sigma(\pi p)/\sigma(pp) \simeq 2/3$. Finally, in the explicit calculation of our amplitude, following \cite{LR}, we indeed found an almost constant slope $B_g \simeq 0.9 ~\GeV^{-2}$ for the present collider energy interval.} $B_g=1$ GeV$^{-2}$.
 To obtain the full opacity we take the integral
\be
\Omega(y)~=~\int_{k^2_t}\Omega(y,k'_t)\frac{dk^{'2}_t}{k^{'2}_t}\ ,
\ee
where the lower limit reflects the $\Theta(k'_t-k_t)$ function in (\ref{lambda}).

Since the opacity $\Omega^a(y,k_t)$ is proportional to $f^a(y,k_t)$ we may write the evolution equation in rapidity $y$, just in terms of opacities.
Thus, finally, we obtain a system of two evolution  equations. 
One equation evolving for $\Omega^b$ up from the target $(b)$ at $y=0$, and one for $\Omega^a$ evolving down from the beam $(a)$ at $y'=Y_k=\ln(s/k^{'2}_t)$. That is
$$\frac{\partial\Omega^b(y,k_t)}{\partial y}=\frac{\alpha_s(k_t)}{2\pi}\int dk^{'2}_t
S(y,y',k_t,k'_t){\cal K}(k_t,k'_t)\Omega^b(y,k'_t)$$
\be
\label{system}
\frac{\partial\Omega^a(y',k_t)}{\partial y'}=\frac{\alpha_s(k_t)}{2\pi}\int dk^{'2}_t
S(y,y',k_t,k'_t){\cal K}(k_t,k'_t)\Omega^a(y',k'_t)
\ee
for the evolution of gluon distributions from both the target and the beam initial hadrons (protons) in the absorptive (background) field of both hadrons. This system can be solved by iteration. In fact, it converges after just a few iterations.

\section{The parton $k_t$ distribution}
 
The transverse momentum distribution at rapidity $y$ has the form
\be
\label{kt-d}
\frac{d\sigma}{dk^2_t}\propto\frac{f^b(y,k_t)f^a(y'=Y-y,k_t)}{k^4_t}\, .
\ee

\begin{figure} 
\begin{center}
\vspace{-5.0cm}
\includegraphics[height=14cm]{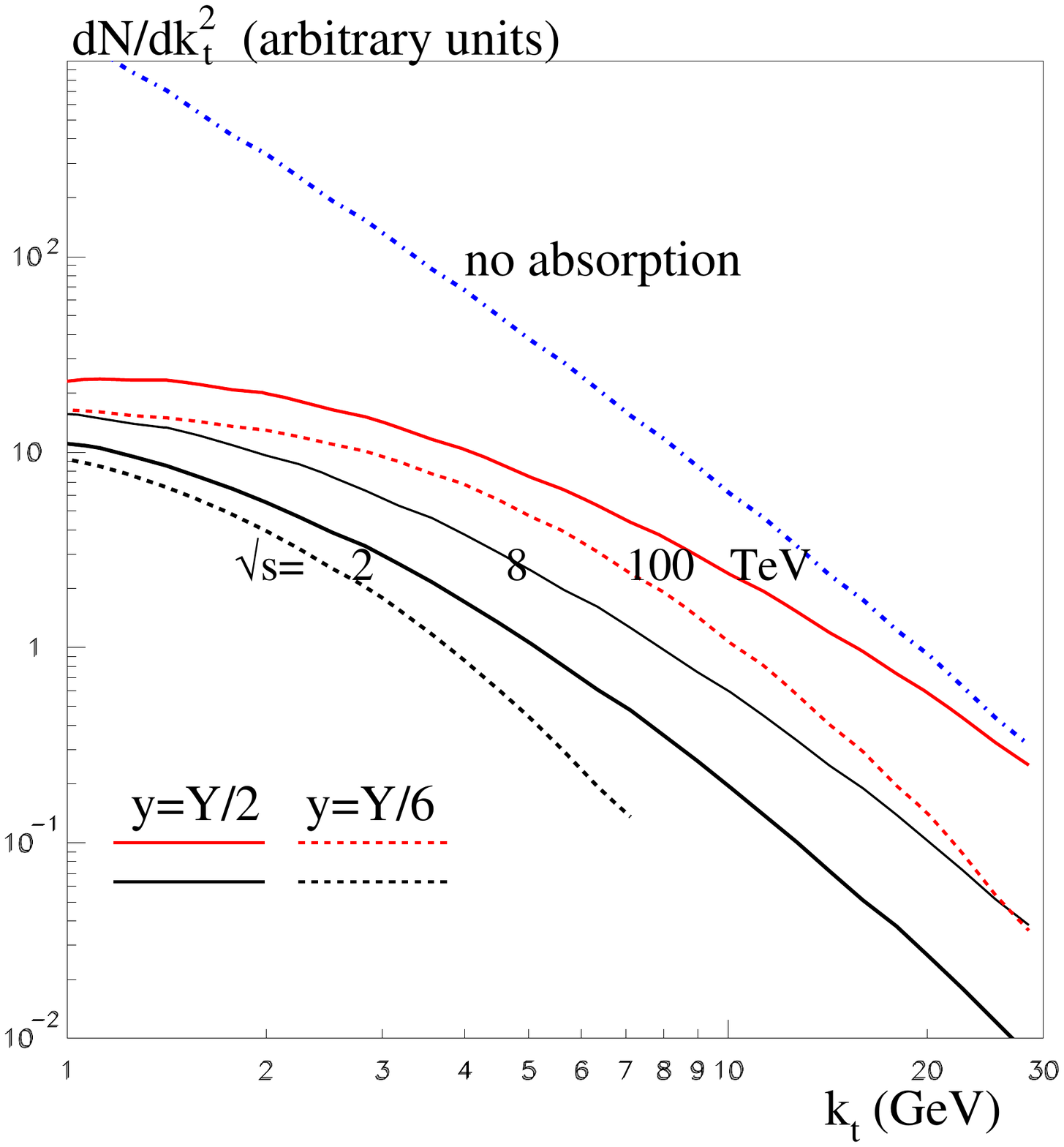}
\caption{\sf The intermediate gluon distribution $dN/dk^2_t$ in the centre of plateau, $y=Y/2$, (solid lines) and near the edge of plateau at $y=Y/6$ (dashed lines, shown for 2 TeV and 100 TeV). The dot-dashed (blue) line shows the distribution
at $\sqrt s=100$ GeV generated if we {\it neglect} the absorptive effects.}
\label{fig:f2}
\end{center}
\end{figure}

The system of equations (\ref{system}) was solved numerically by iteration,
 introducing an infrared cutoff $k_0=0.5$ GeV; that is, assuming 
 $f(y,k_t<k_0)=0$. 
The resulting transverse momentum distributions are presented in Fig.\ref{fig:f2}. The solid lines are the predictions for the gluon distribution in the central plateau region (with rapidity $y=Y/2$), while the dashed lines correspond to distributions shifted to the fragmentation region of the incoming proton (i.e. initial gluon) with $y=Y/6$. The $y=Y/6$ curves are  steeper and the corresponding mean transverse momentum is smaller than that in the centre
($y=Y/2$). As expected the distributions become flatter when energy increases.
However at the Tevatron (thin black curves) and even at the 8 TeV LHC (thick black curve)
 we are still far from true saturation. Only at $\sqrt s=100$ TeV do we predict an horizontal interval for $k_t<2$ GeV. For comparison we present (by a dot-dashed blue line) the distribution expected at $\sqrt s=100$ TeV if one {\em neglects} the absorptive effects, that is for the case when survival factor $S\equiv 1$ in (\ref{system}). The distribution then decreases approximately linearly with increasing $k_t$.

\begin{figure} 
\begin{center}
\vspace{-6.5cm}
\includegraphics[height=14cm]{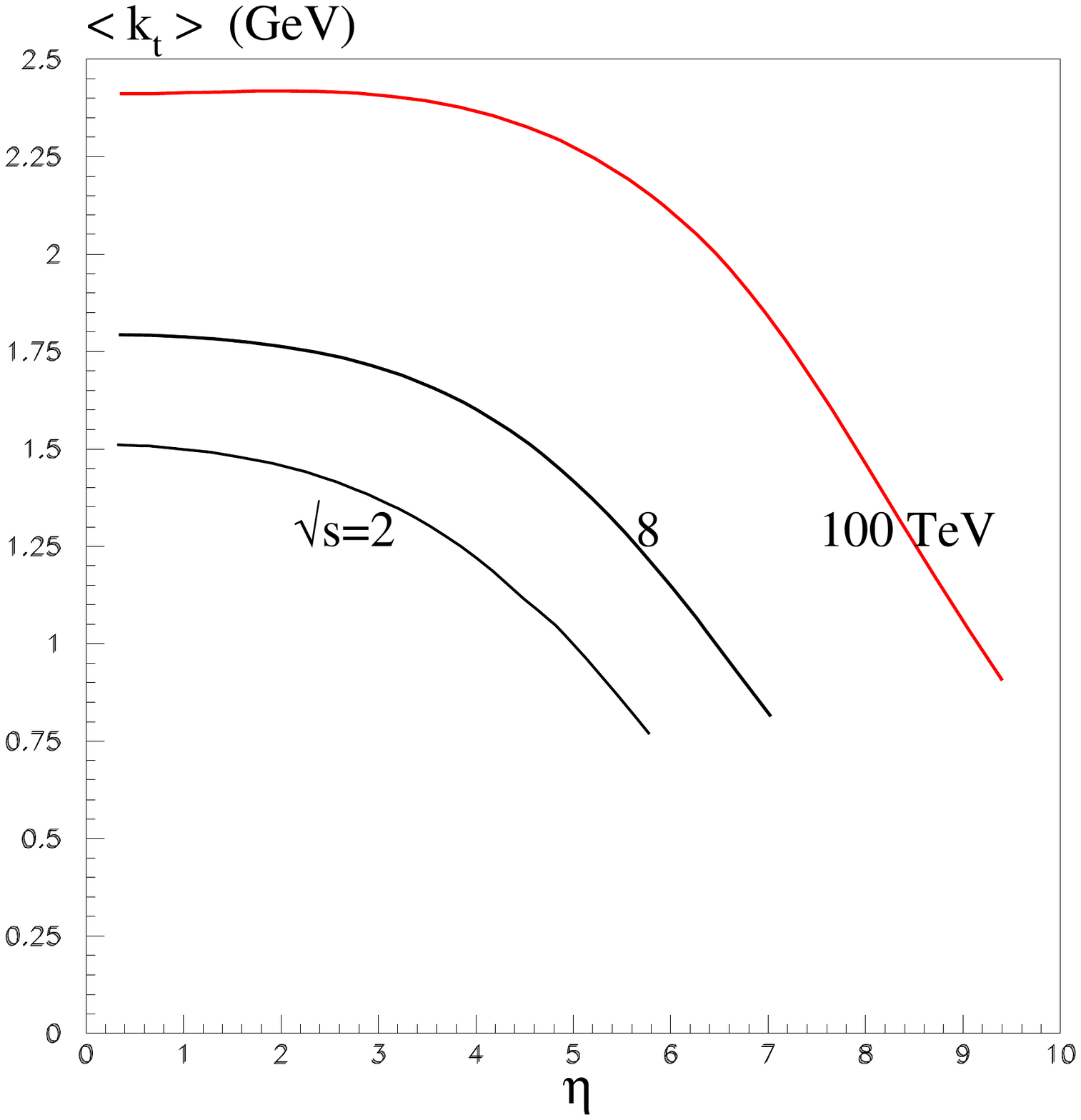}
\caption{\sf The mean transverse momentum, $\langle k_t \rangle$, of the gluon versus the pseudorapidity, $\eta$, of the intermediate gluons.}
\label{fig:f3}
\end{center}
\end{figure}

Next in Fig. \ref{fig:f3} the rapidity dependence of mean transverse momenta, $\langle k_t \rangle$ is shown for the Tevatron ($\sqrt{s}=2$ TeV) and the LHC  ($\sqrt{s}=8$ TeV), and for $\sqrt s=100$
 TeV. The value of mean $\langle k_t\rangle $ increases with energy, and decreases as the 
 rapidity approaches the position of the initial hadron.
 
\section{How to measure the $k_t$ distribution}

Note that the predicted values of $\langle k_t\rangle $ are of the same `order of magnitude', but smaller than, the value of the $k_{\rm min}$ cutoff used in the PYTHIA Monte Carlo, which is based on DGLAP evolution. However, we have to recall that (a) these are not exactly the same quantities and (b) here we have used a simplified model based on the LO BFKL kernel\footnote{Surprisingly, with the same parameters, LO BFKL (supplemented by the simple kinematic constraint and absorptive multi-pomeron effects) leads to an effective gluon-gluon (hot-spot) interaction that increases like $s^{0.15}$ in the present collider energy interval, which is in reasonable agreement with the intercept needed to describe the experimental data.}.  The advantage of this model is that it is sufficiently transparent and practically has no free parameters. The only exceptions are the starting values of $x_0=0.2$ and $k_0=0.5$ GeV and the slope $B_g=1$ GeV$^{-2}$ of the initial `constituent' gluon.  The parameters are not chosen to describe the data, but simply taken to have physically reasonable values.
Besides this, there may be some `intrinsic' transverse momentum of the initial gluon which will enlarge the final value of $\langle k_t\rangle $. 

We should emphasize that the partonic $k_t$ distribution, although not directly observable, drives all soft high energy interactions. Clearly
 it would be interesting to measure the gluon's $\langle k_{g,t}\rangle $ 
 experimentally. Can this be done? The 
 problem is that actually we never observe partons, but only the final 
secondary hadrons, which are mainly pions. Unfortunately the distributions of light hadrons (such as pions, kaons) are strongly affected by final-state interactions: that is by 
hadronization, confinement and the decay of resonances. In particular, the $p_t$ distribution of secondary pions strongly depends on the possible colour re-connection. Therefore it appears better to study the distributions of mesons which contain one heavy quark. Due to the strong leading particle effect \cite{LPE}, the $p_t$ distribution of these mesons is close to that of the heavy quark.  Since heavy quarks are mainly produced by the $gg\to Q\bar Q$ 
subprocess, we may expect that (modulo some smearing due to hadronization when the heavy quark picks up a light antiquark) the mean momentum of such a meson should carry the momentum of the parent gluon. Final-state interactions and resonace decays
do not appreciably distort the $p_t$ distributions of these heavy mesons.

On one hand, it might be the best to measure the $p_t$ distributions of heavy B-mesons, where the leading particle effect is more pronounced. On the other hand, the $b$-quark already receives a rather large 
\be
k_t=k_{\rm background}\sim m_b
\ee
 from the hard $gg\to b\bar b$ subprocess and it may be hard to observe the variation of the incoming gluon $\langle k_{g,t}\rangle$ on the top of this large `background', $k_{\rm background}$. Therefore, it seems better to detect D-mesons where the value of $k_{\rm background}\sim m_c$ is comparable with the expected gluon's $\langle k_{g,t}\rangle$. We would hope to observe the growth of
$\langle p_{D,t}\rangle$ with energy at fixed rapidity, and a decrease of $\langle p_{D,t}\rangle$ with
pseudo-rapidity\footnote{Measured in the laboratory frame ($\eta=-\ln\tan(\theta_{\rm lab}/2)$).} at a fixed energy. The last effect can be observed by a comparison of the CMS/ATLAS data at $\eta=0$ with the LHCb data at $\eta=3$ - 4. 

Moreover, note that it possible to do better.  We could suppress the $k_{\rm background}$ contribution generated into the 'hard' $gg\to Q\bar Q$ subprocess if the transverse momenta of 
both heavy mesons ($D$ and $\bar D$ or $B$ and $\bar B$)  are measured. In such a  case the transverse momentum of the $Q\bar Q$ pair is simply equal to the momentum of the parent gluon pair. Of course, we can not avoid the smearing due to hadronization, but it is not so large since it is controlled by the confinement scale and not by the heavy quark mass. So it would be good to measure the vector sum of the momenta of the two heavy meson, or just the complanarity between the two heavy mesons.
Non-complanarity should increase with energy, but decrease with $\eta$.

Another attractive measurement is to compare the $p_t$ of the secondaries produced in the diffractive dissociation with those from non-diffractive inelastic  events.
 It is usually expected (see, for example, \cite{old}),
that the spectra of particles produced in proton diffractive dissociation into a high mass $(M_X)$ state, are similar to that in normal inelastic events taken at an energy $\sqrt s=M_X$. That is in the situation when the energies of the final states are the same. On the contrary, in the picture described above, even in the case of dissociation the $p_t$ distribution of secondaries should be driven by the parton's
 $k_t$ formed by the whole initial energy $\sqrt s \gg M_X$. 
 That is, it does not matter whether the events have a Large Rapidity Gap (LRG) or not.  One consequence (see, also, \cite{probes}) is that
  in  proton diffractive dissociation to a large $M_X$ system (but still $M_X\ll \sqrt s$) the dissociation events, especially near the edge of the Large Rapidity Gap, are expected to have a larger $p_t$
 than those in a normal inelastic $pp$-collision at $\sqrt s=M_X$; modulo to possible hadronization effects.   Moreover the rapidity dependence of the $p_t$ spectra in LRG events are also similar to that in the inelastic interaction at full proton-proton energy $\sqrt s$, and not to the inelastic events with the  proton-proton energy equal to  $M_X$.
  Again, to reduce the effects of hadronization, it would be better to make the comparison by measuring the 
  distributions of $D$-mesons both in inelastic and high-mass dissociation events.

\section{Conclusions}

The transverse momentum distribution of partons plays a pivotal underlying role both in the spectral shape of secondaries and in the asymptotic behaviour of high energy proton-proton collisions. At first sight, just from dimensional arguments, we expect $d\sigma /dk_t^2 \propto 1/k^4_t$.  That is, the major contribution should come from low $k_t$, close to the cutoff ($\lapproxeq 0.3$ GeV) provided by confinement. On the contrary, to describe the data, it was necessary to introduce a much higher cutoff, $k_{\rm min}$, in the hard matrix element of the order of a few GeV, with a value that increase with collider energy, like $s^{0.12}$. Actually such a $k_{\rm min}$ was obtained by tuning the Monte Carlo generators \cite{Pythia,DGLAPMC}, but clearly it should be of theoretical origin. Moreover, $k_{\rm min}$  of the order of a few GeV should be explained in terms of perturbative QCD.

Here, we use a model based on the LO BFKL equation, supplemented by absorptive multi-pomeron corrections. The original BFKL equation includes diffusion in log$k_t$, with, at each step of the evolution, the possibility that $k_t$ may increase or decrease with equal probabilities. However strong absorption of low $k_t$ partons, leads to a growth of $\langle k_t \rangle$ with collider energy. We demonstrate that this effect naturally explains the observed energy behaviour of the effective cutoff, $k_{\rm min}$.

We did not perform a fit to the data, but show, at a qualitative level, that a simplified model based on leading order perturbative QCD with a few physically motivated parameters, produces a reasonable $k_t$ distribution of the partons. We present the expected $k_t$ distributions at different collider energies, and the dependence of $\langle k_t\rangle$ on the energy and rapidity of the parton.

Although the $k_t$ of the parton is not directly observable, we discuss the possibility to experimentally verify these predictions. One way, is to measure the $p_t$ distributions of mesons containing a heavy $c$ or $b$ quark, or better to measure $D\bar{D}$ or $B\bar{B}$ meson pairs. Another possibility is to compare the $p_t$ spectra of diffractive dissociation events with those of non-diffractive inelastic scattering.

\section*{Acknowledgements}

MGR thanks the IPPP at the University of Durham for hospitality. This work was supported by the  RSCF grant 14-22-00281.

\end{document}